# Integrated All-Optical Fast Fourier Transform: Design and Sensitivity Analysis

HANI NEJADRIAHI,[1] DAVID HILLERKUSS,[2] JONATHAN K. GEORGE,[1] VOLKER J. SORGER[1*]

[1]Deparmtent of Electrical and Computer Engineering, George Washington University, 800 22nd St. Northwest, Washington, D.C. 20052, USA
[2]Huawei Technologies Düsseldorf GmbH, Optical & Quantum Laboratory, Riesstrasse 25-C3, 80992 Munich, Germany
*Corresponding author: sorger@gwu.edu

**The fast Fourier transform (FFT) is a useful and prevalent algorithm in signal processing. It characterizes the spectral components of a signal, or is used in combination with other operations to perform more complex computations such as filtering, convolution, and correlation. Digital FFTs are limited in speed by the necessity of moving charge within logic gates. An analog temporal FFT in fiber optics has been demonstrated with highest data bandwidth. However, the implementation with discrete fiber optic FFT components is bulky. Here, we present and analyze a design of an optical FFT in Silicon photonics and evaluate its performance with respect to variations in phase and amplitude. We discuss the impact of the deployed devices on the FFT's transfer function quality as defined by the transmission output power as a function of frequency, detuning phase, optical delay, and loss.**

The fast Fourier transform (FFT) has applications ranging from signal filtering, to orthogonal frequency division multiplexing (OFDM) transmission [1, 2]. While digital electronic FFT implementations are limited by the physical charging of wires and the necessity of driving logic gates to perform the arithmetic operations digitally, optics naturally enables Fourier transform functionality; for instance a spatial Fourier transform when propagating through a lens [3], and temporal FFT via Cooley-Tukey butterfly pattern [4, 5]. Spatial optical Fourier transforms are generally challenged by bulky free-space optics, however recent advances in meta-surfaces showing flat-lens functionality [6] may enable more compact forms. An efficient temporal FFT has been realized in fiber optics treating the two operations, addition and multiplication, by phase-cascaded delayed interferometers and sampling the output with electro-optic modulators [4]. In theory, such a system is entirely passive, apart from the samplers, and the FFT rate simply depends on the physical propagation delays of the optical signal in the FFT structure. As demonstrated, it allows for wavelength division multiplexing (WDM) showing tens of Tbit/s of processing bandwidth [2].

Here we show a design of an $N = 4$ optical FFT (OFFT) integrated in Silicon photonics for signal modulation frequency of $f_s = 10$ GHz. We explore its performance by analyzing the quality of the transfer function and perform a component-based sensitivity analysis to understand its operation limitations. By relating the physical behavior of the component level to the transfer function, we study the resilience of the OFFTs with respect to fabrication imperfections and thermal effects. These two effects lead to interferometer imbalances due to optical loss variations, phase detuning of the interferometers, and even temporal delay variations. The unreliability in fabrication originates from material selection and processing conditions, wafer dicing, bonding techniques, etc. While absolute delay and losses are fixed after fabrication, the introduced phase uncertainties can be compensated with phase shifter, here through heaters, to ensure the proper functionality of the optical filtering.

The FFT is performed by the Cooley-Turkey method where signal addition is realized by integrated 2x2 directional couplers, while phase multiplication is implemented by a relative phase difference [4, 5]. The resulting butterflies are all passive components, but may require active tuning for phase alignment against drift as discussed below [Fig. 1(a)].

Following the signal path from generation (this could be off-chip) to the OE conversion at the samples, the signal enters the FFT via grating couplers and is 2:1 fanned-out in two stages [Fig. 1(b)]. The signal to be processed can be generated either ON or OFF-chip. Interestingly, the FFT delay decreases linearly with increasing modulation rate, a unique feature of optics. The on-chip portion of the OFFT consists of cascaded delayed interferometers and passive components such as directional couplers, y branches, straight and spiral waveguides, and operates on time domain signals. We opted for the silicon on insulator (SOI) platform due to its high refractive index difference, stable fabrication process, and low cost [7]. While this work is a theoretical analysis of this on-chip OFFT, we briefly discuss our tapeouted process (chip delivery pending), since it the design-to-reality discrepancies are of interest here. All passive design, the on-chip OFFT including all passive components and heaters for phase tunability are fabricated at IME's Silicon Photoncis general-purpose fabricationl. process. An extra oxide cladding is deposited to reduce propagation loss. To take advantage of the thermo-optic effects in silicon photonics, metallization and selective oxide release are selected to create the active components needed for tuning the refractive index of silicon as a function of temperature (resistive heating process).

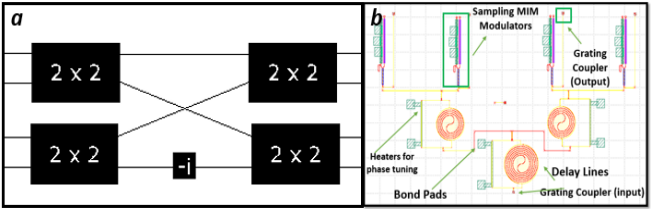

Fig. 1. (a) Cooly-Tukey Butterfly schematic of OFFT (b) Optical $N = 4$ FFT on chip with heaters and modulation on-chip (ideal design) with a waveguide width of 0.5 $\mu m$ and the heaters width were 8 $\mu m$ to enable the heaters carry twice the level of current and the resistance will be about half the previous value. $\lambda_{signal}$ = 1550 nm. OFFT substrate uses a silicon thickness of 220 nm with buried oxide (BOX) thickness of 2 $\mu m$.

The length of the MZI is determined by the system frequency. For the system frequency, $f_s$, the delay, $T$, of waveguides inside the MZIs is determined by $f_s = 10\ GHz$ - where the effective index is: $n_{eff} = 2.5$, the time delay becomes $T_{Delay} = 1/f_s = 10^{-10}s$ and the physical length, $d_{10} = (c/n)T_{delay} = 12\ mm$. For the 10 GHz sampling frequency, the MZIs in the first stage OFFT, with T/2 has a length of 6 mm and for T/4 is 3 mm. The length of the shorter arms of the MZI for the first stage is 500 $\mu m$ and for the second stage is 440 $\mu m$.

The MZI must compensate for phase shifts created by fabrication and temperature variance. We selected the short arm of the MZI heater-tunable over a $\pi/2$ relative phase shift and we can use the following calculation for the needed temperature change. $\frac{dn_{eff}}{dT} \approx \frac{dn}{dT}$ and for Silicon $\frac{dn}{dT} = 1.9 \times 10^{-4} K^{-1}$, giving:
$$\Delta\phi = \frac{\pi}{2} = \frac{2\pi}{\lambda}\frac{dn}{dT}\Delta TL \quad (1)$$

For the second MZI with $L = 500\ \mu m$ the temperature change required (solving for $\Delta T$) is 4.2 K. Thus, we find that temperature is a critical parameter that must be maintained. To do so, we integrate resistive heaters on one of the MZI arms such that with the change in temperature, the desired refractive index change provides the relative phase shift (in this case $\pi/2$) that is required to perform the FFT. Heaters tune effective index of the Silicon waveguide via the known thermo-optic effect. Interestingly, higher modulating rates dictate a smaller waveguide length difference, hence requiring a higher temperature for detuning. Solving for $\Delta T$ we obtain a temperature change of 4.2 K to create a $\pi/2$ phase shift. Thus, heaters are capable of producing the necessary phase shift to compensate for fabrication variance with the appropriate control and feedback.

We selected a minimum waveguide bending radius of 50 $\mu m$ to keep the radiative bending losses low. This results in a delay-line spiral area of 3.9x10$^{-3}$ mm$^2$ for $T$. The total area of this $N = 4$ OFFT is 0.012 mm$^2$. To make this design more compact the sampling in the active design can be done on-chip using MZI modulators following Ref. [8]. This increases the total OFFT area by about 60% to 0.019 mm$^2$.

Sampling is required to obtain the frequency components of the transfer function of the FFT, performed here via electro-optic modulators (EOM). The OFFT design allows placing the EOMs either before or after the FFT butterflies. However, in order to minimize detuning phase, delay, and optical loss differences across the four modulators, the EOMs can appear at the end of the last stage of OFFT to ensure synchronous sampling (Fig. 2). Here we opted for Michelson modulators rather than MZIs due to their smaller (about 50%) $v_\pi L_\pi$ when operated in DC voltages. This is especially rewarding in optical circuitry for minimizing power. To avoid power mismatch loss from the difference in waveguide length of the cascaded MZIs, 'waivy' waveguide bends are added to the shorter arm of the interferometers to compensate for the power loss at the output of the couplers.

The transfer function of the entire system was studied to obtain performance characteristics. The FFT separates frequency contributions of the temporal input signal. Since the system frequency is 10 GHz, the frequency spacing of the OFFT output channels are approximately 10 GHz, but the exact location of the probe frequency for which the maximum transmission is obtained for different outputs is a function of time delay and can vary across the outputs. For every transmission peak at a specific frequency the contributions from the neighboring channels is either minimal

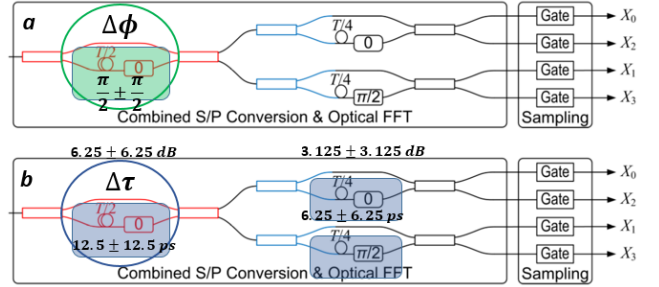

or zero [Fig. 3(a)].

Fig. 2. (a) Sensitivity analysis tests setup of the OFFT's first stage interferometer in terms of phase sweep from π/2 ± π/2 with increments of π/100. (b) Time delay from 12.5 ±12.5 ps (T/4 ±T/4) ps with increments of 0.5 ps corresponding to an optical loss from 6.25 ± 6.25 dB with increments of 0.5 dB for the first stage Mach-Zehnder interferometer.

We are interested in the FFTs performance with detuning critical parameters such as phase, delay, and loss from their design components. Understanding the phase sensitivity and variations across devices can be critical as has been seen during development of ring-resonators [9]. This is obtained by defining performance parameters, and analyzing the transfer function and the extinction ratio of the cascaded interferometers. The discrete nature of the OFFT yields intrinsic quantization errors in frequency and sampling artifacts that are a function of the phase. First, we sweep the phase (0 to $\pi$), time delay and loss in particular that of the lower arm of the interferometer in the first stage of the OFFT (Fig. 2) and analyze the change in the transfer functions of the output. This location is particularly impactful on the transfer function, since it has the highest oscillation and narrow spacing in the frequency domain [5]. Our observable is the frequency detuning of the maximum point of the transfer function, which for instance for output port $X_2$ appears near 6.8 GHz [Fig. 3(a)]. At the ideal case for vanishing phase detuning, the entire power exits to the second branch $X_2$ of the top interferometer due to the additional phase, i.e. the relative phase difference in bidirectional coupler.

Fig. 3. (a) Frequency Sensitivity Analysis on the transmission power (transfer function) of OFFT at ideal phase (b) Phase Sensitivity Analysis on the transmission power at probe frequency of 6.78 GHz.

With phase detuning away from the ideal design point however, the energy of the output transfer functions leaks to the neighboring ports [Fig. 3(b)]; for instance, the power of the first interferometer shifts to the top of the second stage lower interferometer exiting of at the next frequency probe (16.78 GHz), since each output bin in the frequency domain has a 10 GHz spacing given by the system frequency. By sweeping the frequency, we can find points at which the OFFT has full transmission, these points are defined by time delay. Hence the effective impact of the phase detuning is a lateral frequency shift. In the ideal case of no additional phase and at different probe frequencies (frequency at which each OFFT output has a maximum transmission in power), full transmission can be achieved. However sweeping the phase, changes the output amplitudes respectively where the maximum transmission decreases until it vanished and the next full transmission from another output port is achieved, which is a cyclic behavior. Note that the value of 6.78 GHz is not significant as it can be shifted by detuning the delay lines. However, to determine the overall quality of this cascaded system of interferometers, we set a threshold for the phase-detuned power ratio (i.e. frequency power leakage) between target-to-neighboring output ports, while the threshold selection is arbitrary and application dependent, we here chose -20 dB channel-to-channel separation typical for telecommunication applications [5]. Given this threshold, the maximum phase tolerance is < 0.2 radians to ensure acceptable spectral leakage, i.e. channel crosstalk [Fig. 4(a)]. Physically this range corresponds to a small (0.54 K) temperature change that the waveguide index can tolerate to keep within the -20 dB attenuation threshold. Consequently, the phase control must be precise and the required temperature difference needs careful environment control to be achieved. One potential approach is to place the OFFT chip in an ambient chamber with temperature isolation such that the heat could be transported to only the specific areas as desired. Alternatively, control loops and temperature stabilization of the chip could also be employed. Indeed, we observe a non-linear phase error, which is likely due to nature of cascaded interferometers and their phase sensitivity with respect to physical delay lines. To probe the effects of phase detuning errors and distortions in the signal further, we study the difference in the transmission power as a function of phase ($P_{degradation}$) [Fig. 4]. Thus, the $SNR = \frac{(P_{out} - P_{degradation})}{P_{degradation}}$ is obtained by taking the difference in the transmission output power values relative to the ideal zero phase. The SNR indicates the degradation in the system as a function phase detuning and determines the performance quality of the OFFT in response to optical noise created from phase. Here we aim to understand the range at which the OFFT can be operated with minimal sacrifice in power and/or maximum stability and quality. We define the power mismatch ration and figure of merit (FOM)_as follows:

$$P_{mismatchRatio} = \frac{P_{out1}}{P_{out2}}(\phi) \#(2)$$

$$FOM = \frac{SNR}{P_{mismatchRatio}} \#(3)$$

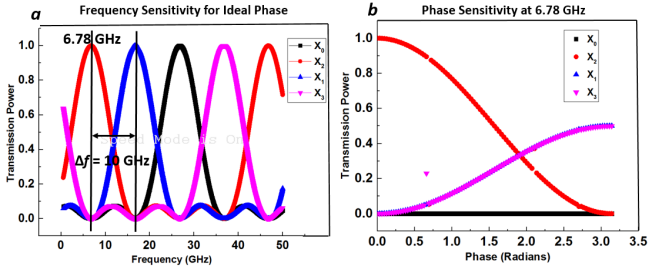

The idea behind the definition of FOM is that the smaller the power mismatch ratio (deviation from unity), and the higher the SNR in the system, the higher is the quality of the OFFT as a function of detuning phase. As a result, the system has the highest FOM at the ideal zero phase case as expected for $X_2$ since the power mismatch ratio between $X_2$ and $X_0$ is close to 1, where the FOM diverges. For the case of $X_1$ and $X_3$ however, their SNR values are low, despite the power mismatch ratio being close to unity 1, since their transmission is minimal for frequency contribution at 6.78 GHz.. The maximum FOM for all four outputs aligns with the design probe frequency value for each frequency bin. As the phase is swept the FOM also decreases drastically for $X_2$ and even further for $X_3$ since the OFFT output is no longer at the probe frequency (max. transmission), as a result SNR decreases as well.

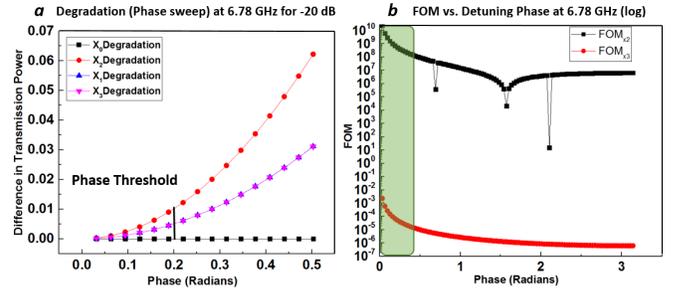

Fig. 4. Degradation for -20 dB tolerance (a) and FOM (b) as a function of detuning phase at 6.78 GHz probe frequency- where $X_2$ has the ideal FOM while $X_3$ is minimal leakage from $X_3$ bin (similar to $X_0$ and $X_1$, not shown).

For perfect OFFT filtering and optimal transmission, the output power of the cascaded MZI arms must match. This however is challenging since in the OFFT design the MZI arms have different physical lengths (waveguides) hence in order to understand how the difference in length changes the quality of the OFFT output, the MZI extinction ratio (ER) was analyzed on the sweeping of the delay/additional loss in the first and correspondingly second stage of OFFT. Shown in [Fig. 2(b)]. The extinction ratio is defined as $ER = \frac{P_{max}}{P_{min}}(\gamma_{loss})$ where $P_{max}$ ($P_{min}$) is the maximum (minimum) power at the output of the OFFT. Delay lines corresponding to loss values were swept across the lower arm of the cascaded MZIs. Note that in the second stage the delay is half of the first stage, and so is the loss. This is important for consistency and symmetry in the overall system design. As proven analytically by the ideal coupler's extinction ratio behavior in the lower arm [Fig. 5(b)], the loss increases exponential with waveguide length as expected. However, the loss increases if the length imbalance increases. This is because of the extra power mismatch between the MZI arms, impacting the quality of the OFFT's transfer function. Thus, the aim is to maximize ER similar to modulators [10] and switches [11, 12], but with the difference of improving the power mismatch between the MZI arms rather signal on-off ratio.

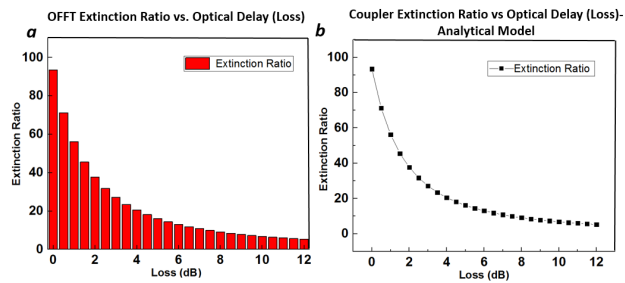

Fig. 5. (a) Extinction ratio of the OFFT full system as a function of physical optical loss from the spiral waveguides (delay lines) (b) Analytical exponential fit (a) based on an ideal coupler.

The FFT data capacity, the number of bits that can be propagated through the system, depends on the modulation type; assuming QAM 256 for a high SNR channel with a bandwidth of 10 GHz the upper bound for bandwidth is 80 Gbps for a single OFFT channel and 320 Gbps for $N = 4$. While we have analyzed the sensitivity and performance for $N = 4$, it is interesting to ask how larger systems scale. Increasing the number of samples ($N$), our OFFT grows with $(N-1)$ cascaded delayed interferometers and $2(N-1)$ couplers. Unlike an electronic FFT, which scales with approximately $5N \log_2 N$, the optical FFT will need to compensate for increasing optical losses with greater optical power [13]. Our FFT scaling analysis shows performance peaks for the OFFT on chip for small $N$ which outperforms an electronic (NVIDIA P100 GPU) for N < 200 [14].

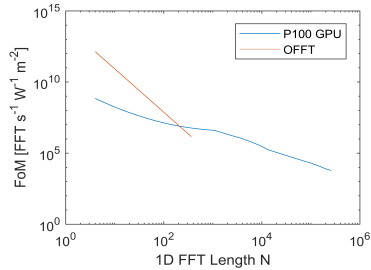

Fig. 6. Analysis of the optical FFT shows up to 3-orders of magnitude higher performance than a GPU (NVIDIA P100) for $N$ < 200 using a figure of merit of 1D FFT per second-Watt-Area assuming insertion loss: 0.9 dB (coupler), 3.5 dB (y-branch), 3.5 dB (modulator), 0.7 dB (first spiral delay line, then with linear scaling), photodetector power of 2.4 µW, and minimum optical power at the photodetector of 250 µW.

In conclusion, we explored the design and operation sensitivity of a temporal $N = 4$ all-optical fast Fourier transform (FFT) on-chip based on silicon photonics. Our design based on cascaded interferometers shows both phase and time delay (loss) sensitivity. We obtain the transfer function of this photonic function by monitoriing the frequency bins at the output ports of the FFT sampled by electro-optic modulators. In our sensitivity analysis we find that the thermal operating range is rather small (<1K) in order to adhere to telecommunication-relevant 0.2 radians phase thresholds with 20 dB tolerance in power loss. Control over this range, however, is possible with thermal on-chip heaters on the MZI arms of the silicon waveguide sections. Unlike electronics, here the number of FFT data processed per second only depends on the time-of-flight of a photon through the millimeter short photonic chip. As such, we find the performance (#FFT data per second, power, and areal footprint) outperforms state-of-the-art graphical processing units GPUs) by 3 orders of magnitude for $N$-scaling below 100. Taken together this temporal FFT shows how photonics enables data processing by simple routing light for an in-the-network-computing, rather than using photons for classical transceiver communication in networks [15].